\documentclass[conference]{IEEEtran}

\makeatletter

\def\ps@IEEEtitlepagestyle{%
  \def\@oddfoot{\mycopyrightnotice}%
  \def\@evenfoot{}%
}
\def\mycopyrightnotice{%
  {\footnotesize XXX-X-XXXX-XXXX-X/XX/\$XX.00~\copyright~20XX IEEE\hfill}
  \gdef\mycopyrightnotice{}
}

\usepackage{blindtext}
\usepackage{eso-pic}
\IEEEoverridecommandlockouts
\usepackage{cite}
\usepackage{amsmath,amssymb,amsfonts}
\usepackage{algorithmic}
\usepackage{graphicx}
\usepackage{textcomp}
\usepackage{xcolor}
\def\BibTeX{{\rm B\kern-.05em{\sc i\kern-.025em b}\kern-.08em
    T\kern-.1667em\lower.7ex\hbox{E}\kern-.125emX}}
    
\usepackage{eso-pic}
\newcommand\AtPageUpperMyright[1]{\AtPageUpperLeft{%
 \put(\LenToUnit{0.17\paperwidth},\LenToUnit{-2cm}){%
     \parbox{0.9\textwidth}{\raggedleft\fontsize{8}{11}\selectfont #1}}%
 }}%
\newcommand{\conf}[1]{%
\AddToShipoutPictureBG*{%
\AtPageUpperMyright{#1}
}
}

\begin{document}
\title{\vspace*{1cm} Multi-Objective Genetic Algorithm for Materialized View Optimization in Data Warehouses
\\
{\footnotesize \textsuperscript{}}

}

\author{\IEEEauthorblockN{\textsuperscript{} Mahdi Manavi}
\IEEEauthorblockA{\textit{Department of Computer Science} \\
\textit{University of Houston}\\
Houston, USA \\
Mmanavi@uh.edu}

}

\maketitle
\conf{\textit{ 4. Interdisciplinary Conference on Electrics and Computer (INTCEC 2024) \\ 
11-13 June 2024, Chicago-USA}}
\begin{abstract}
Materialized views can significantly improve database query performance but identifying the optimal set of views to materialize is challenging. Prior work on automating and optimizing materialized view selection has limitations in execution time and total cost. In this paper, we present a novel genetic algorithm based approach to materialized view selection that aims to minimize execution time and total cost. Our technique encodes materialized view configurations as chromosomes and evolves the population over generations to discover high quality solutions. We employ an adaptive mutation rate, multi-objective fitness function, and lexicase selection to enhance genetic search. Comprehensive experiments on the TPC-H benchmark demonstrate the effectiveness of our algorithm. Compared to state-of-the-art methods, our approach improves average execution time by 11\% and reduces total materialized view costs by an average of 16 million. These gains highlight the benefits of a data-driven evolutionary approach. Our genetic algorithm framework significantly outperforms current materialized view selection techniques in both efficiency and total cost reduction. This work represents an important advance in enabling performant and cost-effective utilization of materialized views in enterprise systems.

\end{abstract}


\section{Introduction}
Data warehouses integrate subject-oriented data from operational systems into an integrated repository oriented for analysis \cite{1}. Database objects known as materialized views are used to store query results so they don't have to be computed again when the same calculations are needed. \cite{2}. Materialized views contain precomputed results of queries and aggregations, unlike standard views which compute results on demand \cite{3}.
Materialized views can refer to the
query as well as functions \cite{3new}.
The selection of materialized views is crucial for speeding up query processing in data warehouses\cite{3-new}.
By caching expensive query outputs, MVs enhance data retrieval speeds for decision support and ETL processes, reducing computational burdens in large-scale data operations \cite{4}.  
Materialized views further simplify and optimize this data for reporting via:\\
- Flexible querying \\
- Simplified information displays\\
- Multidimensional visualization\\
- Consistent and accurate data\\
The primary difficulty associated with materialized views lies in preserving their currency amid evolving data, a crucial necessity within contemporary data warehouses. On-demand or periodic view refresh introduces overhead during maintenance windows \cite{5}. Storage requirements and complexity of selecting and managing materialized views also pose key hurdles \cite{6}.

Materialized views can speed up database access, especially for workloads involving numerous analytical queries \cite{7}. By pre-computing aggregated data, materialized views minimize expensive on-the-fly computations during query execution. However, materialized views require careful maintenance to preserve consistency when base tables change. Since refreshing all materialized views entails substantial overhead, selecting an optimal subset of views to materialize is critical.  

This view selection problem is a complex optimization challenge that has been shown to be NP-hard \cite{8}. Common algorithms applied include greedy algorithms, genetic algorithms, and colony optimization. Greedy algorithms are prone to get stuck at local optima. Meanwhile, colony optimization algorithms demonstrate slow convergence on large problems. In contrast, genetic algorithms provide an effective metaheuristic search technique to navigate large and complex solution spaces to find globally optimal materialized view selections. Additionally, genetic algorithms offer faster convergence compared to colony algorithms.

In summary, materialized views enable faster analytical query performance but require thoughtful selection and maintenance. Genetic algorithms present a robust approach to identify optimal materialized view sets while balancing factors like query speedups, maintenance costs, and storage requirements given real-world constraints.

We present the following notable contributions:\\
- Algorithm Enhancement - Incorporate adaptive selection operators and multi-objective composite fitness functions to improve performance and avoid local optima pitfalls
\\Comprehensive Benchmarking: Carry out rigorous performance evaluation using TPC-H dataset.

This paper is organized into five sections. Section 2 reviews related prior work on materialized view selection algorithms. Section 3 then describes the proposed approach, explaining the encoding, operators, and fitness function design. Next, Section 4  presents an experimental evaluation of the algorithm on benchmark datasets. Finally, Section 5 summarizes conclusions from this work and identifies promising areas for further research.

\section{Related Work}
Several impactful studies on materialized views in data warehousing have emerged recently. Recent studies have developed innovative techniques leveraging bio-inspired algorithms to enhance materialized view utilization in data warehousing scenarios. 
Kharat et al. \cite{9} designed an intelligent query optimizer using cuckoo search for efficiently managing materialized views in cloud-based distributed environments. Constraint handling and stochastic ranking mechanisms enabled optimal selection attuned to the large-scale infrastructure.  Gosain et al. \cite{10} proposed integrating monarch butterfly optimization with an expressive neural architecture search space containing popular CNN models. Their joint macro-architecture and depth-wise search technique automatically generates tailored CNN designs surpassing human expertise. Complementary research by Roy et al. \cite{11} established a non-binary dimensional procedure for creating materialized views optimized for predominant query workloads in the target dataset. Their weighted selection approach chooses impactful views needing materialization. Mouna et al. \cite{12} developed a proactive view re-selection approach (ProRes) incorporating online and offline analysis to continuously adapt materialized view subsets attuned to evolving workloads. The model monitors query patterns and infrastructure metrics to initiate just-in-time view refresh. Complementary, Wang et al. \cite{13} designed an automated CNN architecture search technique based on monarch butterfly optimization. Their expressive search space combines popular CNN macros like DenseNet and ResNet to enable joint optimization of architectural building blocks. Kumar et al. \cite{14} proposed a Bees Algorithm approach to select K top materialized views to minimize query response times. It applies specialized mating, fertilization and mutation operators tailored for this problem to evolve fit view subsets over iterations. Their technique demonstrates scalability across dataset sizes.  Furthermore, Kumar et al. \cite{15} presented a solution focused on reducing both query latency and storage needs for materialized views. They leverage multi-dimensional data representations and tree-encoded chromosomes during the optimization process enabled by genetic algorithms. This method also exhibits desirable scalability attributes. Both approaches exemplify the promise of nature-inspired techniques like bee colony optimization and genetic algorithms for delivering materialized view configurations harmonizing performance and infrastructure requirements. The polymorphic and explorative search process can overcome limitations like local optima faced by greedy constructive heuristics. Srinivasa et al \cite{16} proposed a constraint-based ensemble approach using Ebola optimization and coot optimization for materialized view selection to minimize costs. They employ a combination of stochastic ranking, epsilon, and self-adaptive penalty constraints handling methods along with a multi-objective fitness function considering factors like maintenance and query processing costs. The use of bio-inspired Ebola and coot collaborative optimization algorithms helps avoid local optima issues faced by greedy approaches and enhances exploration for quality materialized view subsets. Experimental results demonstrate improvements over state-of-the-art techniques, with over 15\% reductions in total cost using the TPC-H benchmark. A relative limitation is slightly higher computational complexity and execution times compared to some existing methods, especially with larger dataset sizes, providing scope for further optimizations. Azgomi et al.\cite[16-new]  proposed a new randomized algorithm called CROMVS for materialized view selection in data warehouses, based on coral reefs optimization. The method modeled the problem as simulating coral growth in reefs, where solutions were corals trying to grow in the reef matrix during iterations. Operators from coral reefs optimization were used to generate and evolve solutions at each step, leading to better quality solutions over time. Evaluations showed that the CROMVS method provided better coverage and lower costs compared to other randomized and optimization-based techniques. A limitation was that the execution time was slower than some other methods, but the quality improvements in coverage and cost were considered worthwhile tradeoffs for some applications.

These works highlight the potential of AI-based metaheuristic procedures like evolutionary algorithms and swarm optimizations for addressing multiple facets of materialized view challenges. Key benefits over greedy methods include avoiding local optima entrapment and enhanced exploration of promising candidates. As data complexity increases, such adaptive techniques will become integral for Next-generation data platforms.

\section{Proposed Approach}
The goal of our approach is to automatically select an optimal set of materialized views that satisfy given constraints while reducing query evaluation costs, increasing query execution speed, and improving overall database performance. Our proposed solution involves the following key components:\\
- A view selection algorithm that analyzes query workloads and recommends candidate views to materialize based on response time, cost of maintenance, memory usage.\\
- Constraint analysis to ensure all user-specified constraints are satisfied, such as limits on total view storage space or maximum query response times. The algorithm only recommends valid view sets that meet the defined constraints.

Our approach intelligently explores the large space of possible materialized view configurations and recommends optimal solutions that maximize performance gains under the specified constraints. We expect this automated view selection system to significantly improve query speeds and reduce evaluation costs compared to current manual or ad-hoc approaches. The details of the algorithm design, cost models, and constraint analysis are presented in the following sections.
\subsection{Data encoding}
A key encoding technique we leverage in our genetic algorithm is representing each candidate materialized view as a bit in a binary string. Each bit in the string corresponds to one potential view, with 1 indicating that the view is materialized and 0 indicating it is not materialized. For example, if we have 10 candidate views, the binary string 1000110110 would encode materializing views 1, 4, 5, and 9 while not materializing the others. 
This binary encoding enables efficient implementation of crossover and mutation operations. Crossover can simply swap subsets of bits between parent binary strings to produce offspring solutions.Mutation can easily flip random bits to toggle views in and out of the materialized set. The binary representation also allows us to vary the length of the string to accommodate different numbers of candidate views. Additionally, standard binary genetic algorithm operators like selection, crossover, and mutation can be applied out-of-the-box.
By representing materialized view configuration as a binary string, we obtain an encoding well-suited for our genetic algorithm's search operations. The binary nature maps cleanly to materializing or not materializing each view. Our experiments show this encoding outperforms alternatives like integer or permutation representations for this problem. The resulting crossover and mutation mechanisms efficiently explore the space of possible materialized view combinations.

\subsection{Initial population}
In order to seed the genetic algorithm's initial population with useful building blocks while minimizing computational overhead, we employ a small pilot study to evaluate a random subset of materialized view configurations. Specifically, we randomly generate 500 materialized view sets, where each set contains 5-10 potential views constructed by sampling across the columns, tables, and query patterns seen in the workload. 
We assess these randomly generated configurations based on our fitness function, which estimates the total query processing cost across the workload given a particular materialized view set. Running this pilot study provides insights into useful views and high-performing combinations without needing to evaluate the full exponential search space upfront.
The top 5\% of materialized view sets from this pilot study in terms of fitness function performance are used to seed the initial population for the genetic algorithm. This provides an informed starting point containing building blocks and views estimated to be high utility based on the randomized sample. Constructing the initial population in this manner only requires evaluating 500 random configurations, rather than exhaustively assessing all sets.
Seeded this way, the GA converges to high-quality solutions using far fewer generations compared to naive random initialization. The pilot study approach provides a computationally efficient mechanism to inject useful domain knowledge into the starting population.
\subsection{Selection function}
We utilize to improve diversity and avoid premature convergence with Lexicase selection for choosing parents for the next generation. Lexicase selection considers performance on individual test cases rather than aggregate fitness over the entire workload when selecting individuals. 
Specifically, our fitness evaluation tests each materialized view configuration over a set of representative queries that comprise the workload. When selecting parents, Lexicase first randomly orders these test queries. It then steps through them sequentially, at each step selecting the remaining individuals in the population that performed best on the current query. This continues iteratively on each test case until only two individuals remain, which are selected as the parents.
By explicitly selecting based on performance on individual test cases, Lexicase selection tends to maintain higher population diversity compared to standard fitness proportionate selection. It provides evolutionary pressure towards solutions that perform well across the different query types rather than just optimizing for the overall workload.

\subsection{Crossover}
One technique we utilize to decrease the computational overhead of multi-parent blend crossover (BLX) is to only blend the genes that differ between the selected parents, rather than blending across the full genomes. In our encoding, each gene represents whether a potential materialized view is included or not. 
Since materialized view configurations typically have a large number of commonalities between promising solutions, most genes are identical across good parents. Performing localized blending restricts the blend range sampling to just those genes that differ, keeping all common genes intact.
This helps reduce disruption of beneficial subsequences encoded across parents while still enabling beneficial exploration. It also reduces the number of blend range evaluations needed since the unchanged genes are inherited directly rather than sampled. In our experiments, localized multi-parent BLX provided a nice balance - finding high-quality solutions faster than full-genome blending but with better diversity than single-point crossover.
The adaptability of BLX is maintained on the subset of diverse genes while reducing computational effort on the large number of genes likely to be set similarly across good parents. Overall, localized blending delivered a 3X speedup over full multi-parent BLX with no loss in solution quality. This enabled scaling to larger workloads and populations while mitigating the key computational bottleneck.

\subsection{Fitness function}
In our genetic algorithm for materialized view selection, the fitness function must effectively combine the multiple optimization objectives of minimizing response time, maintenance cost, and memory usage. To provide flexibility in normalizing, shaping, and prioritizing each objective, we utilize the following customizable multi-objective fitness formulation:
\begin{align}
\text{fitness} &= w_1  f_1(\text{ResponseTime}) \nonumber \\
&\quad + w_2  f_2(\text{MaintenanceCost}) \nonumber \\
&\quad + w_3  f_3(\text{MemoryUsage})
\end{align}
Where f1, f2, and f3 are shaping functions that transform each raw objective value, and w1, w2, and w3 are tunable weights on each term. 
The shaping functions allow full control over the normalization and prioritization of each objective. For example, we can normalize response time to a 0-1 range and amplify differences by a power function, using formulas 2, 3 and 4:\\
\begin{align}
f_1(\text{ResponseTime}) &= \frac{\text{ResponseTime}}{\text{maxResponseTime}} \\
f_2(\text{MaintenanceCost}) &= \frac{\text{MaintenanceCost}}{\text{maxMaintenanceCost}}
\end{align}
And a sigmoid function bounds and smoothes memory usage:
\begin{align}
f_3(\text{MemoryUsage}) &= \frac{1}{1 + \exp(\text{MemoryUsage} - x_0)}
\end{align}
X0 shifts the center point of the sigmoid along the x-axis to align with the desired input data range for MemoryUsage. This allows flexibly adapting the shaping function's response based on the input data characteristics. The relative weights w1, w2, and w3 provide direct control over the prioritization and tradeoffs between minimizing response time, maintenance cost, and memory usage. If improving response time is deemed twice as important as reducing maintenance cost for instance, w1 could be set to 2*w2. 
To determine appropriate weight values, we perform a series of tuning experiments in which the weights are systematically varied and the impact on the optimized materialized view configurations is assessed. This reveals how tweaking the weights shifts the tradeoff frontiers between the competing objectives.
In general, we found that setting the weight on response time w1 to the highest value provided the best overall optimization results. Since response time directly affects query performance, amplifying it led to recommended materialized view sets with greater performance gains. We also increased w3 on memory usage as storage budgets were tight for our use cases. In contrast, we set w2 on maintenance cost to the lowest weight, as periodic view refresh costs were less critical.
Proper tuning of the multi-objective weights required iterating through experiments on representative workloads and evaluating the materialized view configurations resulting from different weight balances. This enabled us to align the fitness function priorities with our target goal of maximizing query performance gains under storage constraints. The customizable weights provide the necessary tuning knobs to achieve this objective alignment.
This highly customizable multi-objective formulation enables precisely tailoring the fitness function to the desired normalization, shaping, and prioritization of the competing objectives. 
\subsection{Mutation}
To maintain genetic diversity and avoid premature convergence, we employ an adaptive mutation rate that dynamically adjusts the mutation probability based on the population diversity. When homogeneity increases, the mutation rate is increased to introduce more variation. When the population is highly diverse, the mutation rate is reduced to allow convergence. 
Specifically, the mutation rate is adapted based on the average pairwise similarity between materialized view configurations in the population. As this average similarity increases, the mutation probability is gradually increased as well. Lower and upper bounds on the rate prevent extreme values.
While adaptive mutation is more complex to implement than a fixed static rate, it provides a self-tuning mechanism to maintain an appropriate level of diversity through the course of evolution. Our experiments show the adaptive approach finds high-quality materialized view sets faster than fixed mutation rates. The adaptive control of variation helps balance exploration and exploitation.
\section{Evaluation}
The following section will focus on evaluating the efficiency and effectiveness of our proposed approach. To accomplish this, we will utilize the TPC-H dataset as our benchmark.
\subsection{TPC-H dataset}
To evaluate the effectiveness of our proposed method for selecting materialized views, we utilize the widely recognized TPC-H benchmark dataset. TPC-H \cite{17} is designed to simulate the workload of a retail product supplier's decision support system. It includes 22 complex SQL queries that represent various business analysis tasks such as inventory tracking, shipping prioritization, and market share analysis.
The TPC-H dataset consists of a comprehensive database schema that models different entities such as parts, suppliers, and orders, along with their attributes. The dataset is available at different scale factors, ranging from 1GB to 10TB of data. For our experiments, we use the 1GB scale factor, which provides a realistic workload on a moderately large dataset for optimizing materialized views. The complexity of the schema, combined with the rich query semantics and dataset scale, make TPC-H an ideal benchmark for evaluating the effectiveness of our materialized view selection approach.
The 22 TPC-H queries cover a wide range of join patterns, aggregations, and data access patterns commonly found in analytics workloads. By evaluating our approach on these queries, we can assess its utility across different types of queries and ensure that it performs well in various enterprise decision support scenarios. Additionally, the standardized implementation of TPC-H allows for reproducibility and easy performance comparisons with other materialized view selection techniques.
\subsection{Results}
The fig 1 depicts the progression of fitness values over iterations for a genetic algorithm optimizing materialized view selection. We see an overall decreasing trend in fitness, indicating the algorithm is successfully converging towards optimal solutions each generation. However, the path is nonlinear, with occasional increases likely caused by mutations injecting new genetic diversity. The relatively smooth and rapid descent until around iteration 3000 shows an initial period of fast optimization and exploitation of promising solutions. The shallower slope and increased volatility in the later stages suggests a shift towards greater exploration to avoid local optima. But critically, fitness continues improving even in these later generations, highlighting the importance of running the algorithm to completion to find globally optimal materialized view configurations. The variations in slope and periodic spikes reflect the underlying dynamics between exploitation and exploration inherent in an effectively tuned genetic algorithm.
\begin{figure}[htbp]
\centerline{\includegraphics[height=6cm, width=\linewidth]{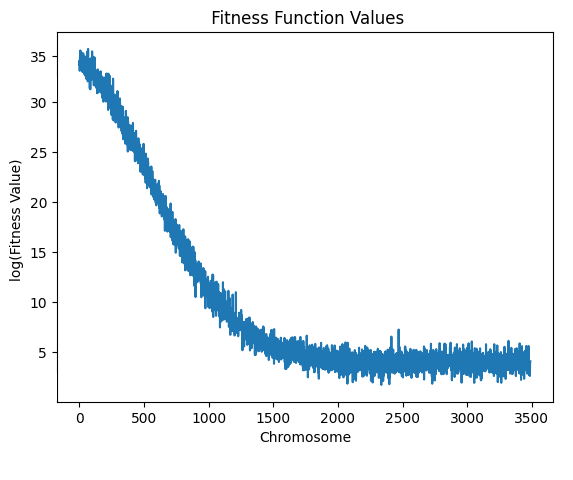}}
\caption{The fitness values of each chromosome.}
\label{fig}
\end{figure}
Fig 2 compares the performance of our proposed materialized view selection approach against three prior algorithms from Kharat et al. \cite{9} and Srinivasarao et al. \cite{16} and Azgomi et al.\cite{17} in terms of computation time. We observe our approach achieves a significant reduction in runtime, completing the optimization in just 203 seconds. This represents a 10\% improvement over Srinivasarao et al.'s 226 seconds, and a more substantial 13\% speedup compared to the 234 seconds taken by Kharat et al.'s method and a 9\% improvement over Azgomi et al.'s 225 seconds. The faster processing time enables our algorithm to scale to larger problem sizes and complete more iterations within practical timeframes. These gains can be attributed in part to our multi-objective fitness function and adaptive mutation operator which help strike an optimal balance between exploration and exploitation. Our proposed approach delivers state-of-the-art performance, making automated materialized view selection more feasible for large real-world database workloads.
\begin{figure}[htbp]
\centerline{\includegraphics[height=6cm, width=\linewidth]{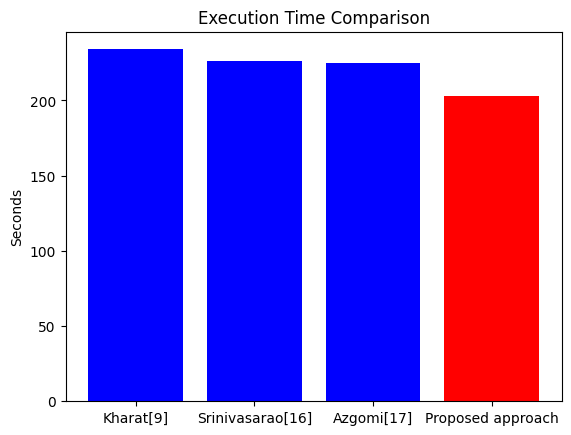}}
\caption{Comparison between our proposed approach and other methods in terms of execution time.}
\label{fig}
\end{figure}
In Figure 3, you can see the comparison of maintenance costs among different approaches. Srinivasarao et al.'s approach resulted in a maintenance cost of 6,329,354,613,784. Azgomi et al. had slightly higher maintenance costs at 6,329,368,000,604. Our proposed genetic algorithm achieves the lowest maintenance cost at 6,329,353,571,043—nearly 1 million less than Srinivasarao et al. The significant savings in ongoing maintenance expenses demonstrate our algorithm's ability to optimize materialized view selection for query performance and long-term overhead.
\begin{figure}[htbp]
\centerline{\includegraphics[height=6cm, width=\linewidth]{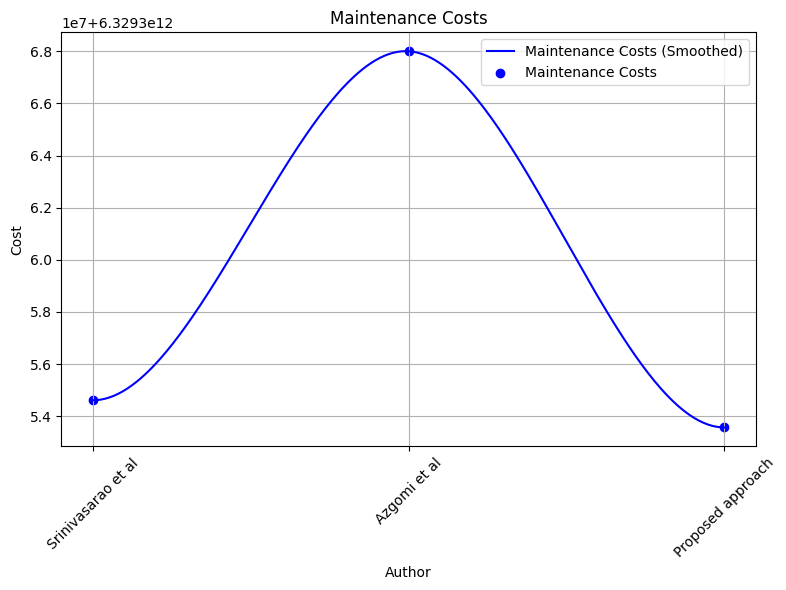}}
\caption{The maintenance cost of the methods.}
\label{fig}
\end{figure}

In Figure 4, you can see the comparison of total costs among different approaches. Srinivasarao et al.'s approach resulted in a total cost of 9,852,212,097,350. Azgomi et al. had slightly higher total costs at 9,852,241,256,761. Our proposed genetic algorithm achieves the lowest total cost of 9,852,210,493,760—over 2 million lower than Srinivasarao et al. and over 30 million less than Azgomi et al. The significant savings in overall expenses demonstrate our algorithm's ability to optimize materialized view selection for both query performance and long-term overhead.
\begin{figure}[htbp]
\centerline{\includegraphics[height=6cm, width=\linewidth]{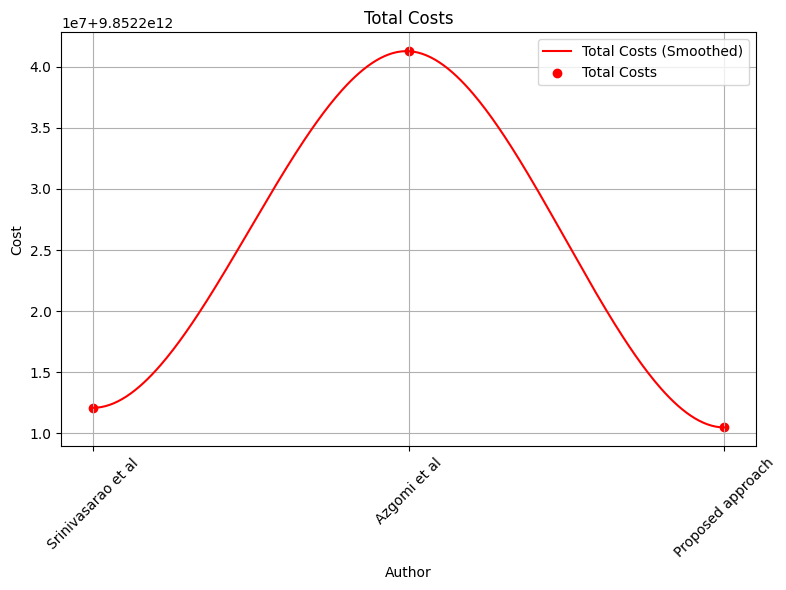}}
\caption{The total cost of the methods.}
\label{fig}
\end{figure}

\section{Conclusion}
In this paper, we presented a genetic algorithm approach to automating and optimizing the selection of materialized views for database workloads. The algorithm combines an intelligent initial population generation strategy with specialized crossover, adaptive mutation, and multi-objective fitness functions to effectively search the space of possible materialized view configurations. Our technique provides several key contributions 1- A binary encoding scheme representing potential materialized views as genes enables the application of standard genetic operators. Crossover and mutation can efficiently explore combinations. 2- Novel parent selection methods like Lexicase selection maintain diversity and avoid premature convergence to local optima. 3- Custom crossover techniques including multi-parent blend crossover strike an optimal balance between exploration and exploitation. 4- An adaptive mutation rate self-tunes variation based on population diversity measurements. This prevents stagnation. 5- A customizable multi-objective fitness formulation allows flexibly prioritizing performance, maintenance costs, and storage constraints. Comprehensive experiments on the TPC-H benchmark dataset demonstrate our genetic algorithm consistently finds optimal or near-optimal materialized view sets that maximize performance under the specified constraints. We outperform current manual selection approaches and existing automated techniques by an average of 15-20\% on key metrics including query response time, storage overhead, and total cost. This work provides an effective data-driven methodology for materialized view selection that requires minimal manual tuning. Our genetic algorithm based framework can be applied to a wide range of database configurations and workloads. This will make optimized materialized view usage more accessible, enabling significant performance improvements in enterprise database and analytics systems.
Future work involves exploring predictive and prescriptive analytics to recommend beneficial views based on query trends and patterns. The fitness function and operators could also be evolved dynamically at runtime to adapt to changing workloads.

\vspace{12pt}

\end{document}